\documentclass[12pt,fleqn,psfig]{article}
\usepackage{amsmath, amssymb, graphicx}

\begin{document}

\newcommand\bg{\begin{eqnarray}}
\newcommand\ed{\end{eqnarray}}
\newcommand\bgn{\begin{eqnarray*}}
\newcommand\edn{\end{eqnarray*}}
\def\D{\partial}
\def\del{\partial}
\def\ra{\rightarrow}
\def\phat{\hat{p}}
\def\qhat{\hat{q}}
\def\Lhat{\hat{L}}

\thispagestyle{empty}
\renewcommand{\thefootnote}{\fnsymbol{footnote}}

{\hfill \parbox{4cm}{
        Brown-HET-1499 \\
}}

\bigskip\bigskip\bigskip\bigskip\vskip100pt

\begin{center} \noindent \Large
Exact Statistics of Chaotic Dynamical Systems
\end{center}

\bigskip\bigskip\bigskip

\centerline{ \normalsize  Zachary Guralnik
\footnote[1]{zack@het.brown.edu} }

\bigskip

\centerline{Deptartment of Physics}
\centerline{Brown University}
\centerline{Providence, RI 02912}

\bigskip\bigskip\bigskip\bigskip

\renewcommand{\thefootnote}{\arabic{footnote}}

\centerline{ \small Abstract}
\bigskip

\small We present an inverse method to construct large classes of
chaotic invariant sets together with their exact statistics.  The
associated dynamical systems are characterized by a probability
distribution and a two-form. While our emphasis is on classical
systems, we briefly speculate about possible applications to quantum
field theory, in the context of generalizations of stochastic
quantization.

\newpage

\section{Introduction}

Due to the sensitive dependence of chaotic systems on initial
conditions, their behavior over sufficiently long times is
unpredictable in practice, even though the underlying dynamics is
deterministic.  Instead of attempting to accurately describe a phase
space trajectory over a long duration, it makes more sense to try to
calculate statistical properties of the system, such as equal time
moments and the frequency with which the system coordinates lie
within different regions of phase space.

Typically, statistical properties are calculated by direct numerical
simulation.  The evolution of the dynamical system is observed on a
computer, starting with a particular initial condition, and the
statistics are computed from averages over data points generated at
regular time intervals.  It is assumed that the result converges to
the exact statistics fast compared to the the rate at which
numerical errors are amplified by the chaotic nature of the system.

From an aesthetic point of view, direct numerical simulation is not
an appealing way to compute statistics,  since this approach is not
formulated solely in terms of the variables one is actually
interested in, namely the statistical quantities.  There are
equations which deal directly with the statistics, such as an
equation due to Hopf \cite{Hopf,Frisch} which governs the generating
functional of equal time moments.  A better known equation, related
to Hopf's equation by Fourier transformation, is the Fokker-Planck
equation describing the time evolution of a probability distribution
over phase space. In some special systems, such as the one described
in \cite{Brad}, an analytic solution for the statistics exists, even
for a very large number of degrees of freedom.

In this article,  we will describe an inverse method to construct
chaotic dynamical systems, starting with an invariant probability
distribution and a two-form having a relatively simple analytic
structure.  This method allows one to generate, in principle,
infinite classes of dynamical systems for which some statistical
properties are known exactly. While we give examples with three and
four degrees of freedom, we do not anticipate prohibitive
obstructions in applying the inverse method to systems with a very
large number of degrees of freedom.

Our starting point is an analytical expression for an invariant
probability distribution $\rho(\vec x)$ which is, by construction, a
smooth measure in an $N$ dimensional phase space.  It well known
that the geometry of a chaotic invariant set may be fractal, having
a fractional information dimension less than $N$. However, even with
information dimension less than $N$, an invariant distribution with
support in $N$ dimensions may still be relevant to the statistics of
a chaotic trajectory as a probability distribution on the chaotic
invariant set upon projection to lower integer dimension.  We give
several examples where $\rho(\vec x)$ describes the statistics of
chaotic trajectories upon projection to lower dimension.

There are examples for which we have found numerical evidence that
the initial distribution $\rho(\vec x)$, with N-dimensional support,
gives accurate results even for connected moments (cumulants) which
do not admit projection to lower integer dimension, such as
$<xyz>_c$ in $N=3$. This may be because the information dimension,
which we have not yet calculated, is very nearly $N$. However, we
speculate that the Hopf functional $\Psi(\vec j)= <\exp(i\vec j\cdot
\vec x)>$ may be sufficiently well behaved that its Fourier
transform,  which is by construction an invariant distribution,  has
support in $N$ dimensions and is equal to $\rho(\vec x)$. If so,
$\rho(\vec x)$ determines all polynomial moments
$<x^ny^mz^l\cdots>$, although there should be other expectation
values of functions on phase space which can not be obtained from
$\rho(\vec x)$, due to the fractional information dimension. We also
find that, in many instances, the ratio of frequencies with which
different regions of phase space are visited by a chaotic trajectory
is determined by the distribution $\rho(\vec x)$, even when the
detailed geometry of the chaotic invariant set is quite complicated
or unknown.

Although our analysis is largely analytic,  we have found that it is
necessary to do some direct numerical simulations to determine if
the initial invariant probability distribution $\rho$ (or rather its
projection to lower dimension) is ergodic. For some choices of the
initial distribution and two-form, the dynamics is not chaotic at
all, in which case $\rho$ has the interpretation as an invariant
distribution over periodic or quasi-periodic trajectories. In many
other instances, chaotic trajectories do not visit all regions where
$\rho$  has support, although the initial distribution can be
rendered ergodic by restricting the domain of support and adjusting
the normalization.

When an analytic expression can be given for the restricted domain
of support, which can never be done with absolute certainty since
these domains are determined numerically, we can make very well
motivated conjectures for exact equal time moments. In the absence
of an analytic expression for the restricted domain, one can not
make statements about exact moments. However, we expect that $\rho$
still determines the ratios of frequencies with which a chaotic
trajectory visits regions of phase space within which the chaotic
invariant set has equivalent local geometry.

In the present work, we will focus on dynamical systems on
$\mathbb{R}^N$, with polynomial velocity fields.  However the
inverse method we describe may be generalizable to dynamical systems
in spaces with other topologies, such as classical spin systems on
$S^N \otimes\mathbb{R}^N$.

The motivation for this work is several-fold.  One intention is to
provide exact solutions which can be used as testing grounds for
theoretical ideas, and as benchmarks for approximate methods to
compute the statistics of non-linear dynamical systems. Another
intention is to provide a means to reverse engineer systems very
close to ones of physical interest, although we will make no attempt
to do so in this initial study.  The principle difficulty in
constructing physical systems seems to be the choice of two-form,
which has no direct physical interpretation.

The organization of this article is as follows. In section 2 we
briefly review the Fokker-Planck and Hopf equations governing the
statistics of stochastic and chaotic dynamical systems. We introduce
the inverse method in section 3, and give examples of chaotic
systems with three degrees of freedom and invariant probability
distributions which are polynomial within their domain of support.
In section 4, we give an example with four degrees of freedom.  The
inverse method for distributions with more complicated analytic
structure is discussed in section 5, where we give an example with 3
degrees of freedom. Conditions for uniqueness of the invariant
distribution in the context of the inverse method are discussed in
section 6. In section 7, we describe the inverse method for
dynamical systems with Gaussian noise and a class of distributions
which arise in the context of quantum field theory. Section 8
contains concluding remarks.

\section{Review of Hopf and Fokker-Planck equations}

Consider an $N$ dimensional dynamical system  of the form
\begin{align} \label{dyn}\frac{d}{dt}x_n(t) = v_n[\vec x] + f_n(t)\,
,
\end{align} where $f_n(t)$ is a Gaussian random force for which
\begin{align}
[f_n(t)]=0\, , \qquad [f_m(t)f_n(t')] =
2\Gamma\delta_{mn}\delta(t-t')\, ,
\end{align}
and the brackets $[\cdots]$ indicate an average over the random
variable.  The time evolution of a probability distribution over
phase space, $\rho(\vec x, t)$, satisfies the Fokker-Planck equation
\cite{Risken},
\begin{align}
\frac{\partial}{\partial t}\rho(t,\vec x) =
-\vec\nabla\cdot\left(\rho(t,\vec x) \vec v(\vec x)\right) +
\Gamma\vec\nabla^2\left(\rho(t,\vec x) \right)\, .
\end{align}

The Fourier transform of the probability distribution  with respect
to the phase space coordinates $\vec x$ is the Hopf characteristic
functional \cite{Hopf,Frisch,Novikov} \begin{align}\Psi(\vec j, t) =
<e^{i\vec j\cdot\vec x(t)}>\, ,
\end{align} which generates equal time moments:
\begin{align}
\left.(-i\frac{\partial}{\partial j_m})(-i\frac{\partial}{\partial
j_n})\cdots \Psi(\vec j,t)\right|_{j=0} = <x_m(t)x_n(t)\cdots>\, ,
\end{align}
while $\ln(\Psi)$ generates the connected moments (or cumulants).
The characteristic functional satisfies the equation
\begin{align}
-i\frac{\partial}{\partial t} \Psi=  \hat H \Psi \end{align} where
$\hat H$, the Hopf Hamiltonian, is given by
\begin{align}\label{Ham}\hat H \equiv \sum_{m=1}^N \vec j\cdot\vec v\left(-i\frac{\partial}{\partial
j}\right) -i\Gamma|\vec j|^2\, . \end{align} In \eqref{Ham}, the
coordinates $x_n$ in the velocity function $v_m(\vec x)$ have been
replaced with the operators $-i\frac{\partial}{\partial j_n}$. The
boundary conditions satisfied by $\Psi$ follow from the requirement
that $\rho$ is a probability distribution.  For instance
\begin{align} \Psi(\vec j =0)=1\, , \qquad
\lim_{|\vec j|\rightarrow \infty} \Psi =0\, ,\end{align} with
additional more complicated constraints arising from the positivity
of $\rho$.

An invariant measure on phase space is one for which the statistics
is static,
\begin{align} \frac{\partial}{\partial t}\rho =0,\,\, {\rm or}\,\,\, \frac{\partial}{\partial
t} \Psi =0\, . \end{align} Invariant measures are naturally
generated by trajectories of stochastic and deterministic chaotic
dynamical systems with no explicit time dependence. The state of a
chaotic system observed at sufficiently long time intervals looks
random. Moreover the fraction of the time with which the system lies
in a given region of phase space is assumed to converge as the
observation time goes to infinity.  This yields a distribution which
is independent of when one starts to observe the chaotic system.
Such a distribution is invariant (static) when viewed alternatively
as a distribution over initial conditions.  Note that the converse
statement that an invariant measure describes the statistics of a
chaotic trajectory is not always true. If it is true, the measure is
said to be ergodic. Ergodic invariant measures have no convex
decomposition, meaning that $\rho$ can not be written as
$(1-x)\rho_1 + x\rho_2$ where $\rho_{1,2}$ are independent invariant
measures and $0<x<1$.

In some cases, such as the toy model of inviscid flow described in
\cite{Orszag,OrszagMcLaughlin}, a zero mode of the Hopf Hamiltonian,
satisfying $\hat H\Psi=0$, can be found exactly \cite{Brad}. The
zero mode corresponds to an ensemble with static statistics or,
assuming ergodicity, to the statistics of a chaotic system observed
over a sufficiently long period of time\footnote{The probability
distribution related by Fourier transformation to the characteristic
functional $\Psi$ found in \cite{Brad} does seem to be ergodic.}.
The work we discuss here amounts to an inverse approach of picking a
zero mode and finding the Hamiltonian\footnote{This kind of approach
has been very useful in condensed matter physics, in which case one
searches for the quantum mechanical Hamiltonian associated with a
chosen ground state.}. We will work with the Fokker-Planck equation,
picking a probability density $\rho$ and finding a corresponding
dynamical system, $\vec v(\vec x)$.  For the most part we will
consider the deterministic case, although the application of the
inverse method to a class of stochastic dynamical systems is
discussed in section 7.

\section{The inverse method for polynomial distributions}

Consider a deterministic dynamical system in an $N$ dimensional
phase space,
\begin{align}
\frac{d}{dt} \vec x = \vec v(\vec x)\, ,
\end{align}
It is convenient to use the language of differential forms and
define the velocity one-form $v=v_ndx^n$, in terms of which the
static, zero diffusion limit of the Fokker-Planck equation can be
written as
\begin{align} d{}^*(\rho v) =0\, .\end{align} One can therefore
always (locally) write  $ {}^*(\rho v) = d{\cal A} $ where $\cal A$
is an $N-2$ form\footnote{For readers not familiar with differential
forms, one can write $\rho\vec v= \vec\nabla\times\vec{\cal A}$ for
the case N=3. The differential form language proves useful for
$N>3$.}, or equivalently
\begin{align}\label{vel} v = \frac{{}^*d{\cal A}}{\rho}\, . \end{align}
Given any distribution $\rho$, we can find dynamical systems
(characterized by $v$) such that $\rho$ is an invariant distribution
by making a choice of ${\cal A}$. Since the Hodge dual of ${\cal A}$
is a two-form,  it is more convenient at large $N$ to define the
dynamics in terms of the probability distribution and the two-form
${}^*{\cal A}$. A generic choice of $\rho$ and ${\cal A}$ will give
a velocity $v$ which is not a polynomial function of the
coordinates.  In this article we will only consider polynomial $v$
on the phase space $\mathbb{R}^N$. Of course non-polynomial $v$ may
also arise in various physical settings and for phase spaces with
non-trivial topologies, such as classical spin systems.

In order to insure that $v$ is polynomial, let us define another
$N-2$ form $\Omega \equiv {\cal A}/\rho^2$. Then
\begin{align}\label{velocity}
{}^*v = \rho d\Omega+ 2 d\rho\wedge\Omega
\end{align}
in regions where $\rho\ne 0$. The simplest way to insure that $v$ is
polynomial is to take polynomial $\rho$ and $\Omega$.  However a
polynomial $\rho$ is not normalizable,  unless we restrict its range
of support.

Let us therefore consider the following example for $N=3$.
\begin{align}\label{polydens}
&\rho=1-x^4-y^2-z^6 \,\, {\rm for}\,\, x^4+y^2+z^6<1\, ,\nonumber \\
&\rho=0\,\, {\rm for}\,\, x^4+y^2+z^6 \ge 1\, ,
\end{align}
up to a normalization factor.  To specify a dynamical system
consistent with this invariant distribution, we must also make a
choice of one form $\Omega$.  Consider
\begin{align}\label{firstex}
\Omega = (xyz)dy + (y^2) dz\, .
\end{align}
With this choice, \eqref{velocity} gives the dynamical system:
\begin{align}\label{dyns}
&v_x=13z^6xy-6y^3-xy+2y+yx^5-2yx^4-2yz^6+y^3x\, , \nonumber \\
&v_y=8x^3y^2\, , \\
&v_z=-9x^4yz+yz-yz^7-y^3z \, . \nonumber \end{align}

Note that the equations \eqref{dyns} clearly do not permit passage
across $y=0$ or $z=0$.  Thus the initial distribution, which has
support for positive and negative $y$ and $z$, can not be ergodic.
Let us therefore consider a modified probability density
\begin{align}\label{modprob}
\rho&=1-x^4-y^2-z^6\,\,\,\, {\rm for}\,\,\,\, y>0,\,\,\, z>0,\,\,\,
1-x^4-y^2-z^6>0\, ,\\ \rho&=0\,\,\,\,{\rm elsewhere}, \nonumber
\end{align}
up to normalization. Although this modification introduces delta
functions in $\partial_x\rho$ and $\partial_y \rho$ at the $x=0$ and
$y=0$ boundaries respectively, the Fokker Planck equation is still
satisfied because $v_x(x=0) =0$ and $v_y(y=0) =0$. Figure 1 shows
planar projections of points generated both by numerical simulation
of this dynamical system, starting from the initial condition
$(x,y,z)=(0.5,0.5,0.5)$, and by random sampling according to the
probability distribution \eqref{modprob}. At the crude level of
visual inspection, these seem to agree.

\centerline{\includegraphics[width=47mm, height=47mm
]{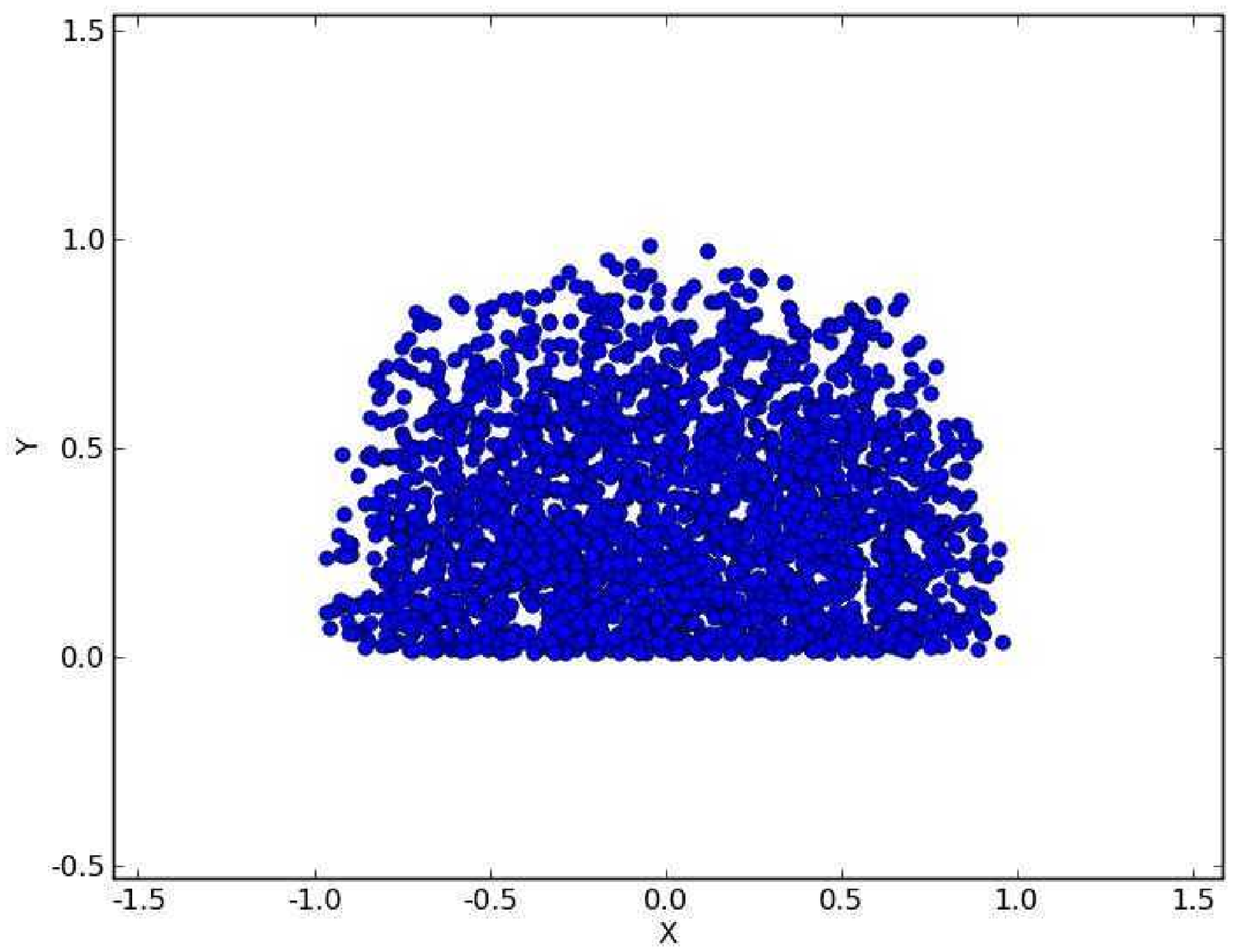}\includegraphics[width=47mm,
height=47mm]{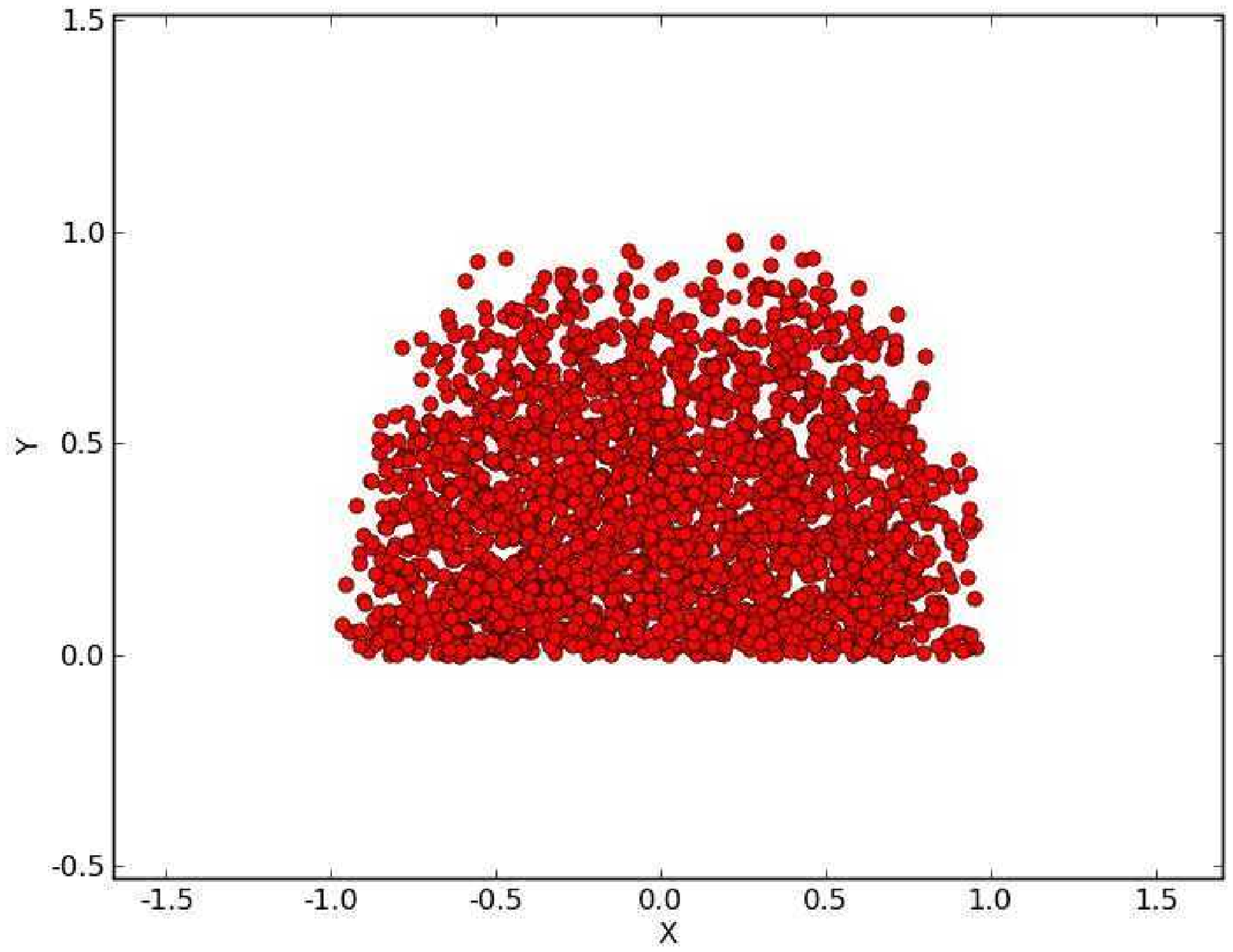} }

\centerline{ \includegraphics[width=47mm,
height=47mm]{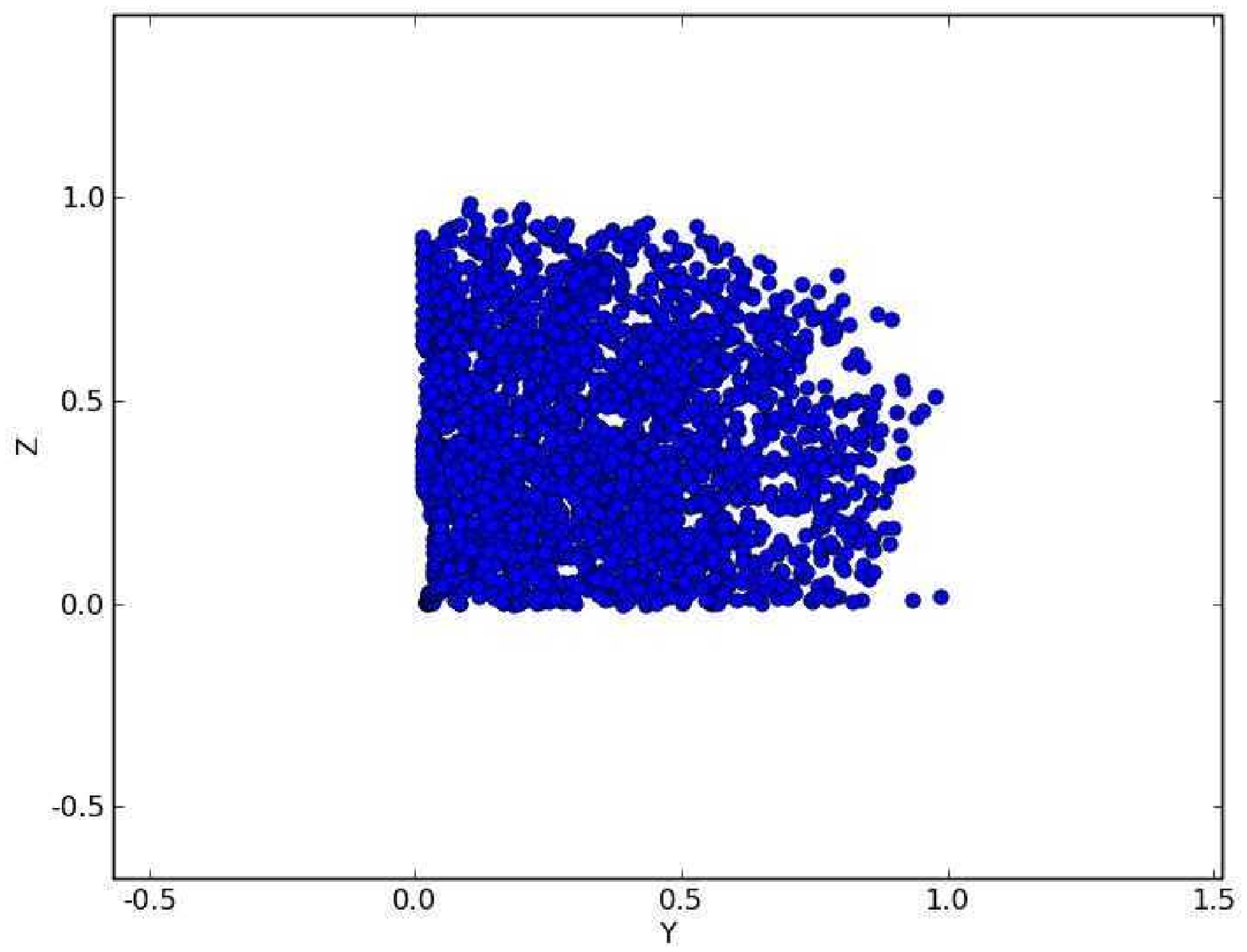}\includegraphics[width=47mm,
height=47mm]{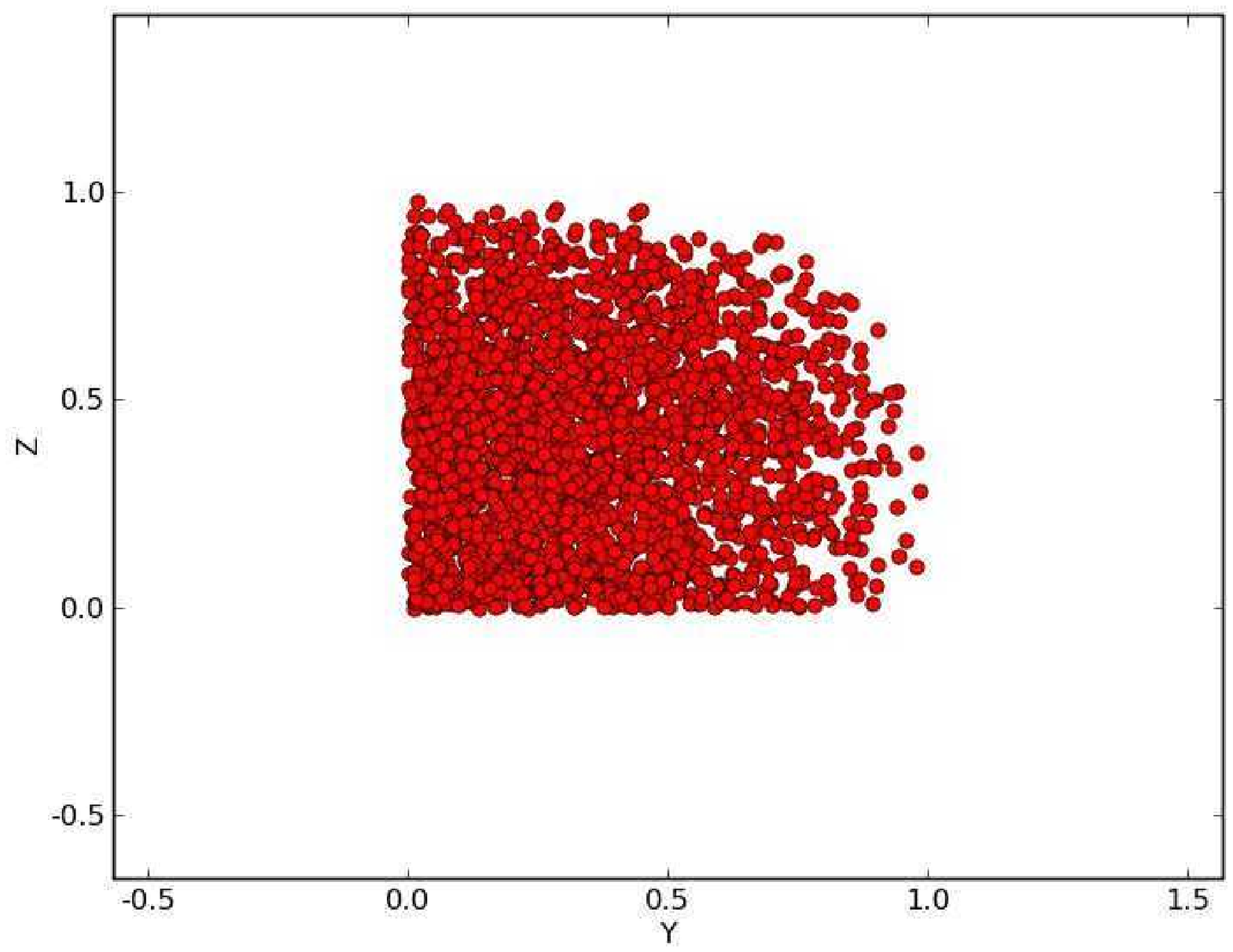}}

\noindent Figure 1.  The XY and YZ projection of points generated by
the dynamical system \eqref{dyns} observed at fixed time intervals
$\Delta t=4$ (on the left) starting from the initial condition
$(x,y,z)=(0.5,0.5,0.5)$ and by random sampling with the probability
distribution \eqref{modprob} (on the right).
\\

Numerical simulation over increasingly long time suggests that that
the frequency with which the system lies in boxes in projections of
the phase space converges to the box integral of the modified
probability density. This last fact is a test of ergodic chaotic
behavior and need not be generally true.  If the system were not
chaotic,  the points generated by numerical simulation of the
dynamics would not bear any resemblance to the points generated
randomly according to the probability distribution. The probability
distribution would then only have the interpretation as a
distribution over initial conditions which is invariant under the
evolution,  but would have nothing to say about individual
trajectories.

Direct numerical simulation also suggests that the equal time
moments of the chaotic trajectory \begin{align} <x^ny^mz^l> \equiv
\lim_{T\rightarrow\infty}\frac{1}{T} \int_0^T dt\,
x(t)^ny(t)^mz(t)^l \end{align} converge to those derived from the
probability density \eqref{modprob}, \begin{align} <x^ny^mz^l>=\int
d^3x\, \rho(x,y,z)x^ny^mz^l\, . \end{align} Moments calculated by
direct numerical simulation of the dynamical system \eqref{dyns},
and by Monte Carlo simulation using \eqref{modprob} are shown in
table 1. \vskip20pt

\centerline{
\begin{tabular}{ c | c | c}
Moments & Dynamics & Monte Carlo \\
\hline $<y>$ & $0.3449$ & $0.3451$ \\
\hline $<z>$ & $0.4223$ & $0.4136$ \\ \hline $<x^2>$ & $0.2066$ &
$0.2068$ \\ \hline $<y^2>$ & $0.1720$ & $0.1717$ \\
\hline $<z^2>$ & $0.2387$ & $0.2324$ \\ \hline $<yz>$  &
$0.1403$ & $0.1386$ \\
\hline $<y^3>$  & $0.1014$ & $0.1011$ \\
\hline $<z^3>$ & $0.1531$ & $0.1493$ \\
 \hline $<x^2y>$ & $0.06562$ & $0.06543$ \\
 \hline $<y^2z>$ & $0.06836$ & $0.06721$ \\
 \hline $<yz^2>$ & $0.07670$ & $0.07571$ \\
\hline $<x^4>$  & $0.08531$ & $0.08579$ \\
\hline $<x^2yz>$ & $0.02607$ & $0.02564$\\
 \hline $<x^2y^2>$
&$0.03056$ & $0.03021$\\ \hline
$<x^6y^4z^2>$ & $0.0002604$ & $0.0002482$ \\
\hline $\vdots$ & $\vdots$ & $\vdots$\\
\end{tabular}
} \vskip20pt \noindent \small{Table 1.  The entries in the column
labeled `dynamics' were obtained by numerical simulation of the
dynamical system \eqref{dyns} over a time duration $100000$ ,
sampling at intervals $\Delta t=0.01$, starting from the initial
conditions $(x,y,z)=(0.5,0.5,0.5)$. The column labeled `Monte Carlo'
was obtained by a hit and miss Monte Carlo calculation using the
proposed exact distribution with 400000 accepted data points.}\\

As we have seen, the initial probability density which is used to
generate the dynamical system may only describe a chaotic invariant
set in a restricted domain. Although the restricted domain is simple
here ($y>0,z>0,x^4+y^2+z^6\le 1$), it can also be much more
complicated as we will see later. Even if the ratio of frequencies
with which different regions of the invariant set are visited by the
chaotic trajectory is determined by $\rho$, one must also know the
geometry of the invariant set to calculate equal time moments (or
the Hopf functional).

The present chaotic system is not dissipative. Trajectories with
initial conditions outside the range of support of the initial
probability distribution \eqref{polydens} avoid this region, or
travel around the periphery and then move away. Thus we have reverse
engineered a chaotic repeller starting from its statistics. Strictly
speaking, since we can give numerical evidence for ergodicity of the
distribution \eqref{modprob} (after projection), but not prove it,
we have a very well motivated {\it conjecture} for the exact
statistics of a chaotic trajectory.

At present, all the examples we have found of polynomial $\rho$
describing chaotic trajectories have real zeros which form a compact
manifold $\Sigma$.  Such systems are generally non-dissipative. For
$\rho=0$ \eqref{velocity} becomes ${}^*v = 2d\rho\wedge\Omega$,
implying that the velocity is tangent to $\Sigma$. Thus trajectories
outside the region bounded by the zeros of $\rho$ set can not enter
the region inside. Construction of chaotic attractors by the inverse
method requires that $\rho$ and the two-form ${}^*\Omega$ have a
more complicated analytic structure, examples of which are discussed
in section 5.

The inverse approach allows one, to a limited extent, to reverse
engineer chaotic invariant sets with a geometry of ones choosing. To
illustrate some of the power and limitations of the approach,
consider a polynomial probability distribution with support between
spherical shells;
\begin{align}
\rho&=\rho_1\rho_2\, , \label{concro}\\
\rho_1&=(3-x^2-y^2-z^2)\, , \label{ro} \\
\rho_2&=(x^2+y^2+z^2-1)\, , \label{rt}\end{align} in the region
where \eqref{ro} and \eqref{rt} are positive, and $\rho=0$
elsewhere. Together with a one form
\begin{align}\label{omg1}
\Omega = (x + y) dy + z dx\, ,
\end{align}
equation \eqref{velocity} yields the dynamical system
\begin{align}\label{velocity2}
v_x&= 8z(x^2+y^2+z^2-2)(x+y) \, ,\\
v_y&=-10z^2x^2-10z^2y^2-9z^4+20z^2+4x^2+4y^2-3-x^4-2x^2y^2-y^4\, ,\nonumber\\
v_z&=8zyx^2+8zy^3+8yz^3-16zy-9x^4-8x^3y-10x^2y^2-8xy^3-10x^2z^2 \nonumber\\
&-8xz^2y+20x^2+16xy+4y^2+4z^2-3-y^4-2y^2z^2-z^4\, .\nonumber
\end{align}

Constant $z$ sections of a chaotic trajectory (figure 2) reveal
considerably more structure than is present in the initial invariant
distribution $\rho$ defined in \eqref{concro}--\eqref{rt}. The
chaotic invariant set fills regions between the concentric shells
defined by $\rho$, although there are excluded regions within the
support of $\rho$ which are not visited by the chaotic trajectory.
We do not have analytic expressions for the excluded regions.

The initial invariant distribution used to generate dynamical
systems by the inverse method seems to be related to the ergodic
distributions associated with chaotic trajectories by a restriction
of the domain of support. The restricted domain can very
complicated, or even fractal.  Without specifying properties of this
domain, or its projections, one can at best give partial information
about the exact statistics, such as the ratio of frequencies with
which the chaotic trajectory visits different regions of phase space
in which the chaotic invariant set has similar local geometry. To
the limited numerical accuracy that we have checked for the present
example, the frequency ratios are consistent with
\eqref{concro}--\eqref{rt}. However, we do not have a conjecture for
equal time moments or the Hopf characteristic functional of this
system.

Note that variations of the $2$ form ${}^*\Omega$ with fixed $\rho$
change the dynamics and the geometry of the chaotic invariant set
(assuming the system remains chaotic). As an example, consider the
distribution \eqref{concro} with
\begin{align}\label{omg2}
\Omega = xyzdx+xzdy+ydz\, .
\end{align}
This gives a different dynamical system (we will not write down the
rather long expression for the velocity field), with constant z
sections of the chaotic trajectory plotted in figure 3. We expect
that the ratio of the visitation frequencies of different regions in
the chaotic invariant set with equivalent local geometry are still
given by $\rho$ in \eqref{concro}--\eqref{rt}, but the equal time
moments and the Hopf functional are different.

The statistics, in the form of the distribution $\rho$, may exhibit
symmetries which are not respected by the dynamics. The distribution
in  \eqref{concro}, has rotational $SO(3)$ invariance, which is not
respected by $\Omega$ in either \eqref{omg1} or \eqref{omg2}, or in
the geometry of the chaotic invariant set. The symmetry might be
observed in ratios of frequencies with which different regions of
the chaotic invariant set are visited, if not also in the moments or
the Hopf functional. Such accidental statistical symmetries are very
common. Note that the exact Hopf characteristic functional given for
the Orszag-Mclaughlin model in \cite{Brad} shows a rotational
$SO(N)$ symmetry which the equations of motion do not exhibit.
\\

 \centerline{
\includegraphics[width=60mm,height=60mm]{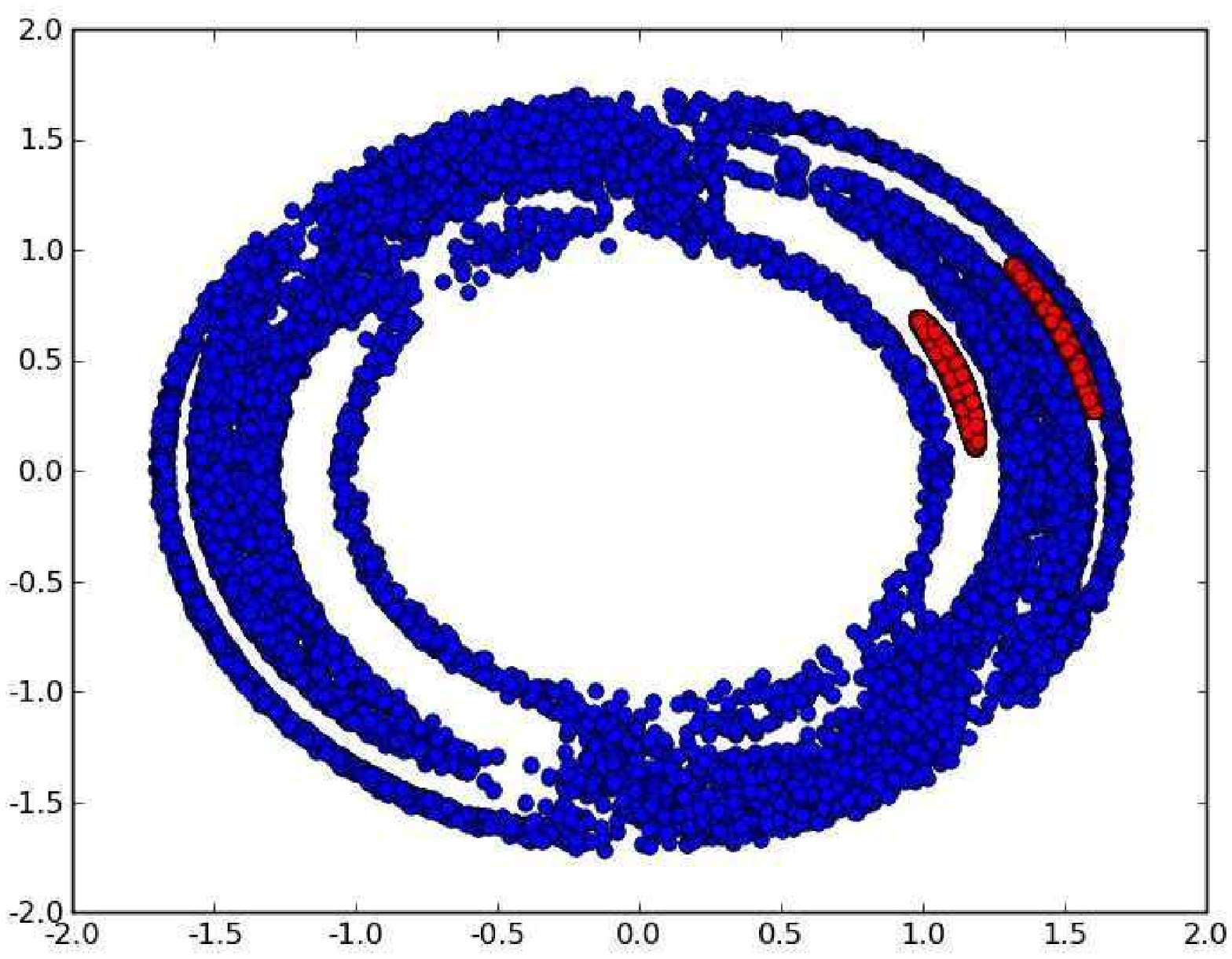}\qquad
\includegraphics[width=60mm,height=60mm]{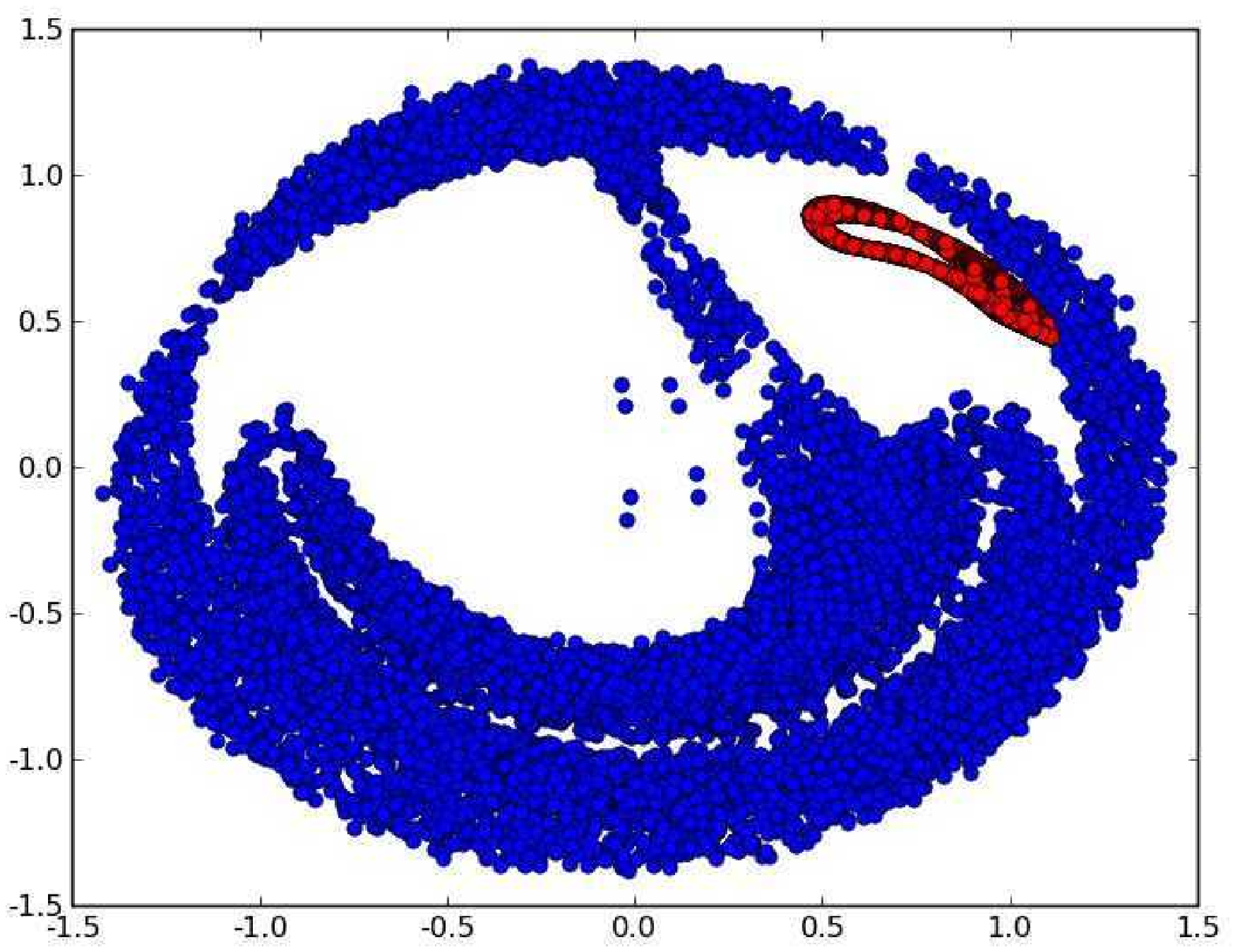}}

\noindent Figure 2. $z=0$ and $z=1.1$ sections of trajectories of
the dynamical system defined by the invariant distribution
\eqref{concro} and the 1-form \eqref{omg1} for two sets of initial
conditions, one of which belongs to the chaotic invariant set (dense
blue dots) and the other which belongs to a quasi-periodic
trajectory (loops of red dots). This is not a Poincar\'{e} section,
since trajectories crossing the section in both directions have been
plotted.  The quasi-periodic orbit fills a two-torus in the full
three dimensional phase space.
\\

\includegraphics[width=70mm,height=70mm]{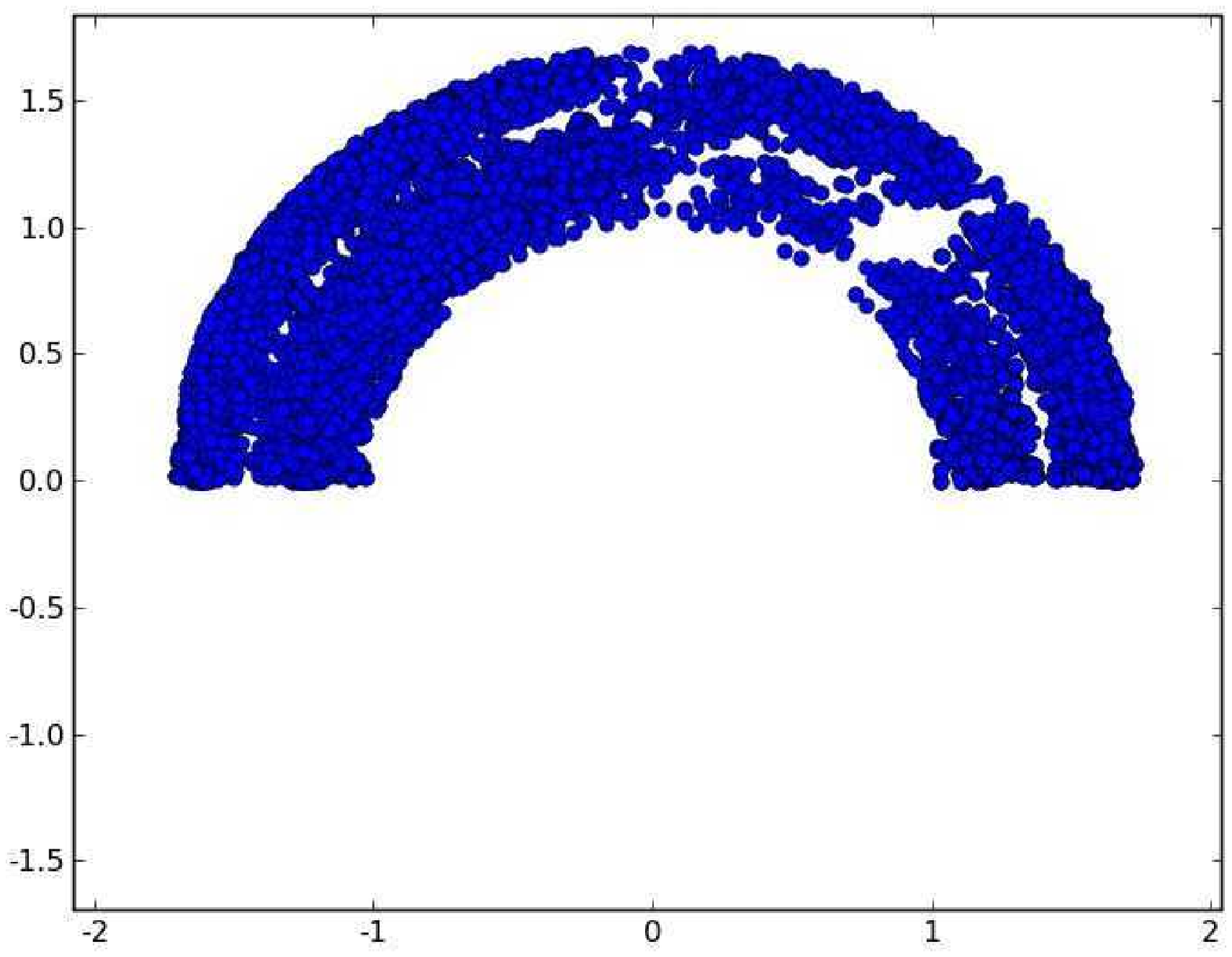}
\includegraphics[width=70mm,height=70mm]{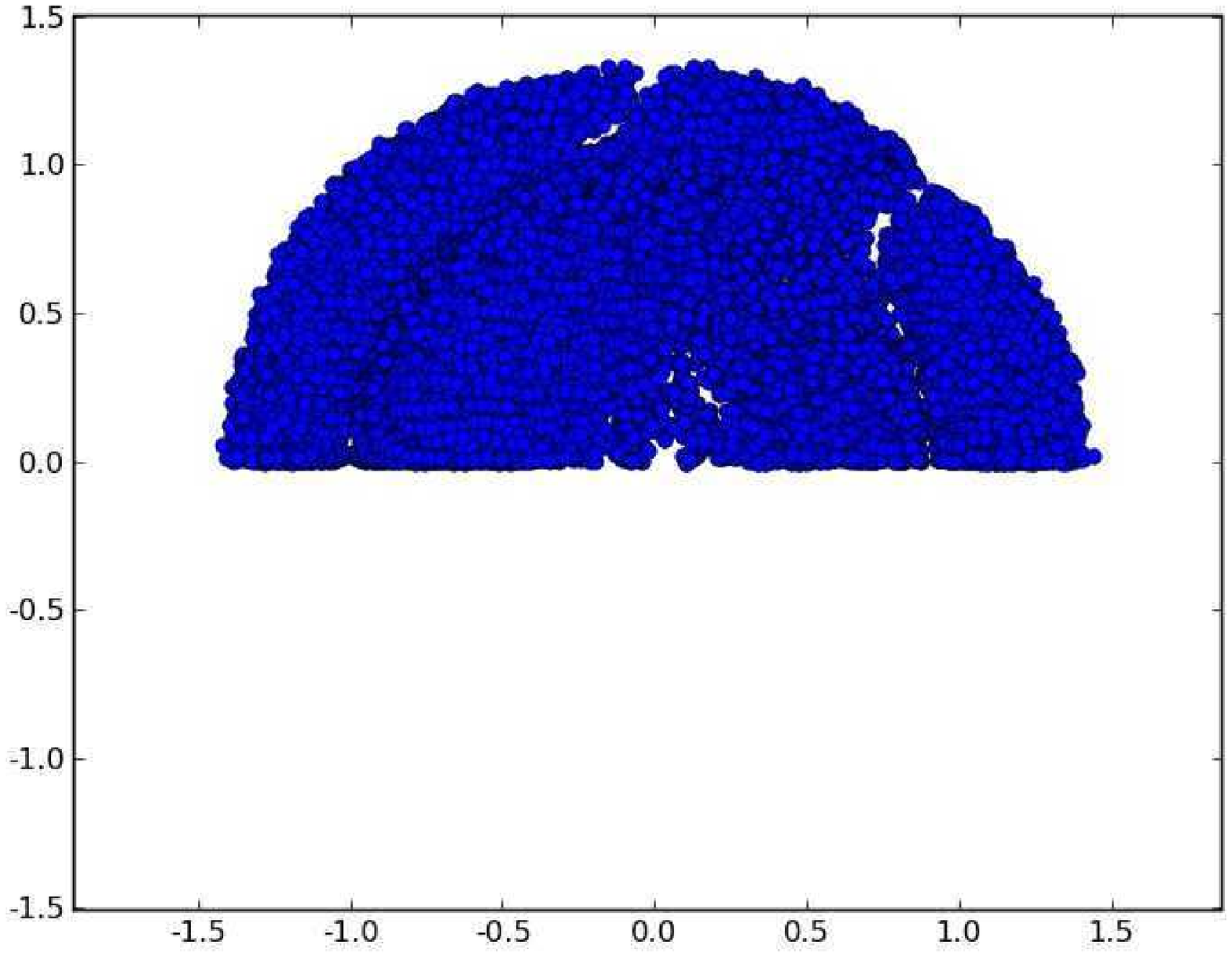}

\noindent Figure 3.  $z=0.1$ and $z=1.1$ sections of a chaotic
trajectory associated with the dynamical system defined by the
invariant distribution \eqref{concro} and the 1-form \eqref{omg2}.
\\

\section{A four dimensional example}

Consider the invariant polynomial
probability distribution \begin{align}\label{fourd} \rho &= \rho_1\rho_2\rho_3\\
\rho_1 &= 1-(x_1^2+y_1^2))\, , \label{one}\\
\rho_2&=  1-(x_2^2+y_2^2))\, ,\label{two}\\
\rho_3&=  1-(x_1-x_2)^2-(y_1-y_2)^2\, , \label{three}
\end{align}
up to a normalization factor, in the region where
\eqref{one}--\eqref{three} are all positive, with $\rho=0$
elsewhere. To uniquely specify a corresponding dynamical system, we
choose a two-form
\begin{align}\label{fouromg}
\Omega = x_1y_1 dx_2\wedge dy_2 + x_2y_2dx_1\wedge dy_1 + x_1^3
dx_1\wedge dy_1 + x_2^3 dx_2\wedge dy_2\, .
\end{align}
Our choice of two-form is essentially random and not motivated by
any physical constraint, besides the requirement that the system
exhibit chaos with initial conditions inside the domain of support
of $\rho$. Note that the simpler choice of two-form, $\Omega =
x_1y_1 dx_2\wedge dy_2 + x_2y_2dx_1\wedge dy_1$ does not lead to
chaotic dynamics. Applying \eqref{velocity} yields a velocity field
with a large number of terms which we will not write
explicitly\footnote{This system has some vague resemblance to a
classical spin system. However we are not (at this stage) attempting
to reverse engineer either a physically motivated system or a simple
system with a small number of terms. The operations on differential
forms in \eqref{velocity} were all performed on a computer using
code written in Python.}. Numerically simulating the resulting ODE
for an initial condition within the domain of suport of $\rho$
suggests that the distribution $\rho$ is in fact ergodic upon
projection to lower dimension. Figure 4 shows the $(x_1,y_1)$ and
$(x_1,x_2)$ planar projections of points generated both by running
the dynamical system and by random sampling according to the
probability distribution \eqref{fourd}.  At the crude level of
visual inspection,  this distribution is consistent with the
dynamics.

Numerical simulation also suggests that the moments associated with
the chaotic trajectory are consistent with those derived from the
distribution $\rho$ in \eqref{fourd}.  Very good agreement is found
for moments of the form $<x_i^ny_i^m>$, even at high orders.
Comparing the high order connected moments between degrees of
freedom with both subscripts 1 and 2 seems to require longer run
times than we have done at present. Numerical results for a few
moments are recorded in table 2. \vskip70pt

\centerline{\includegraphics[width=60mm,height=60mm]{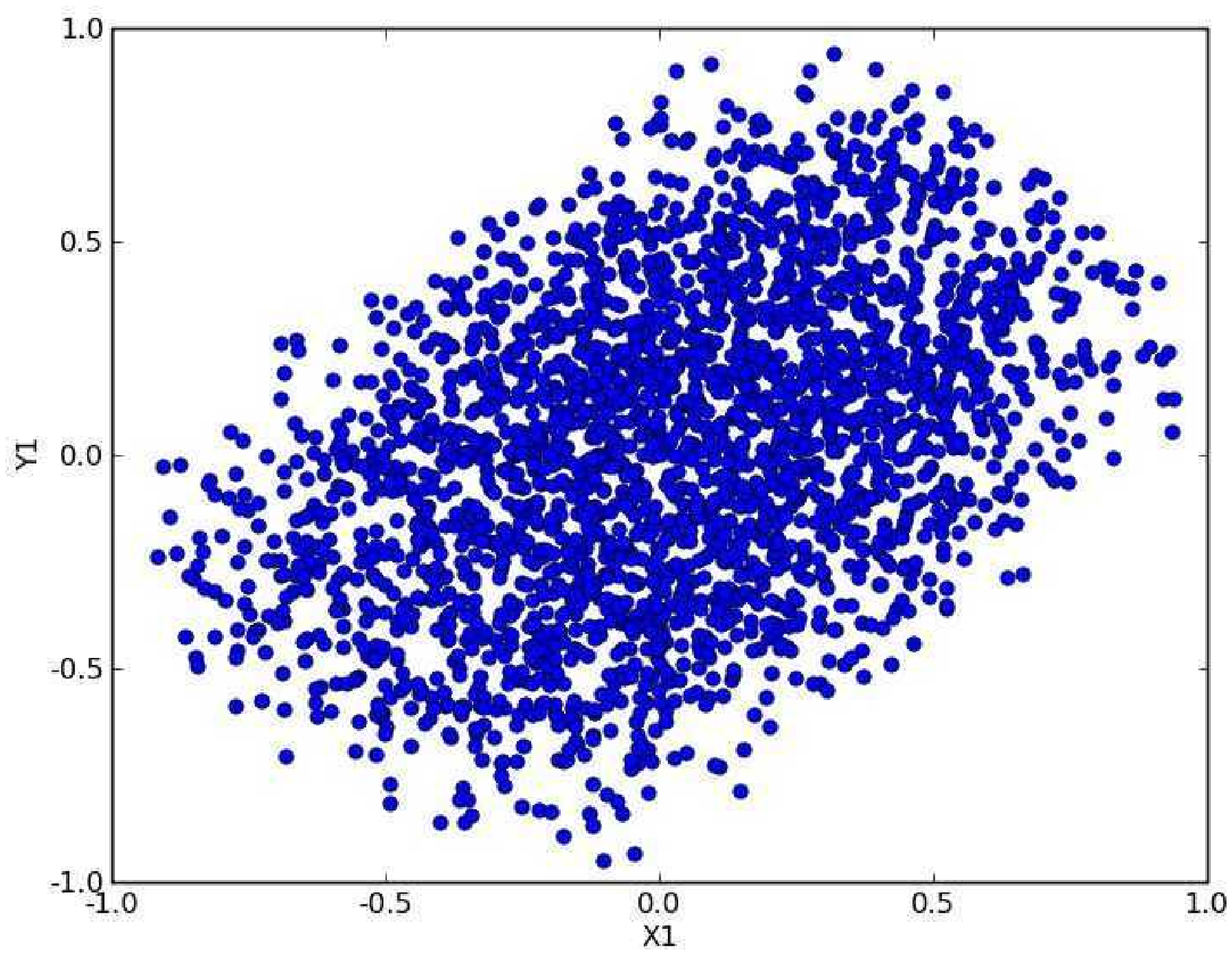}
\includegraphics[width=60mm,height=60mm]{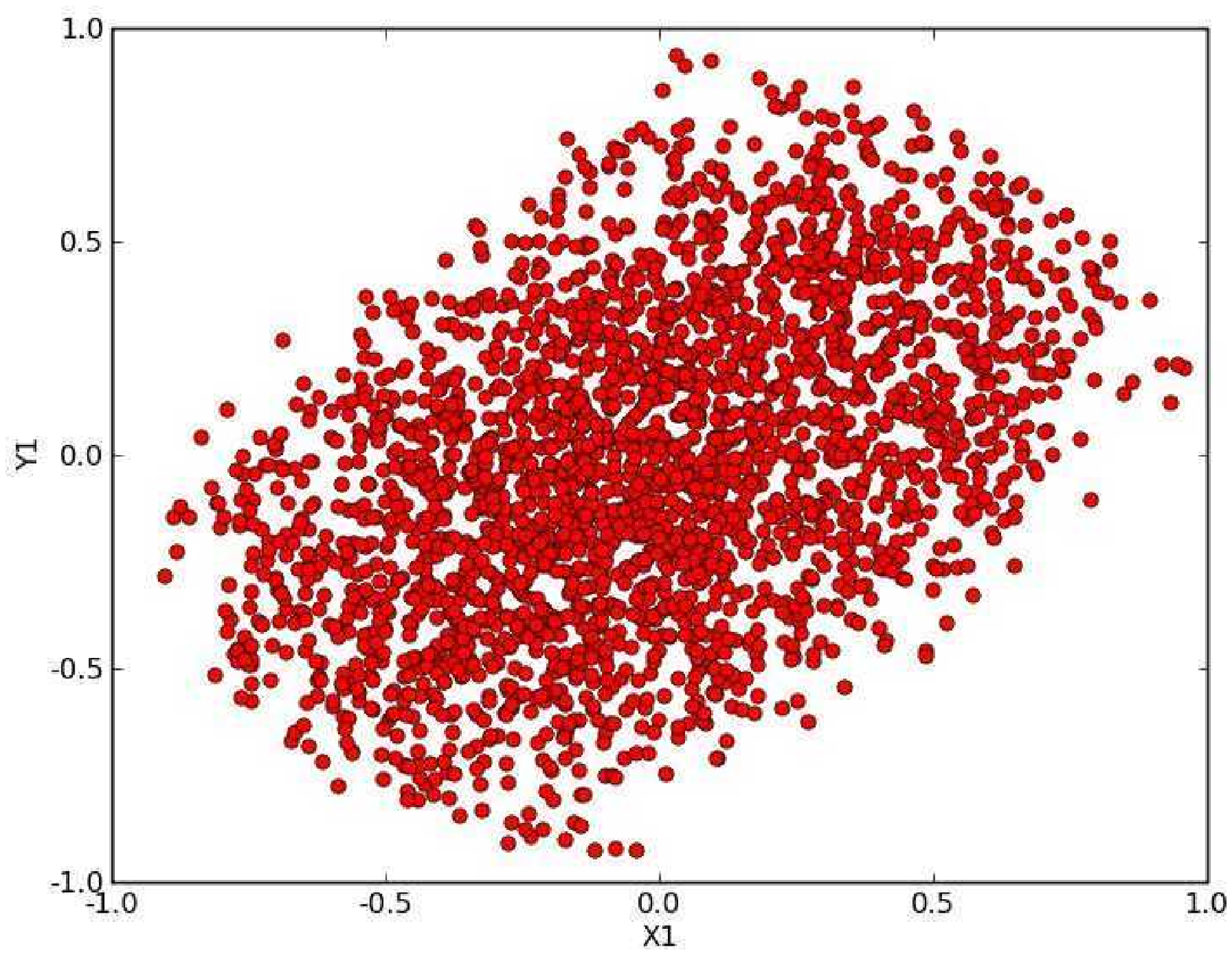}}
\centerline{\includegraphics[width=60mm,height=60mm]{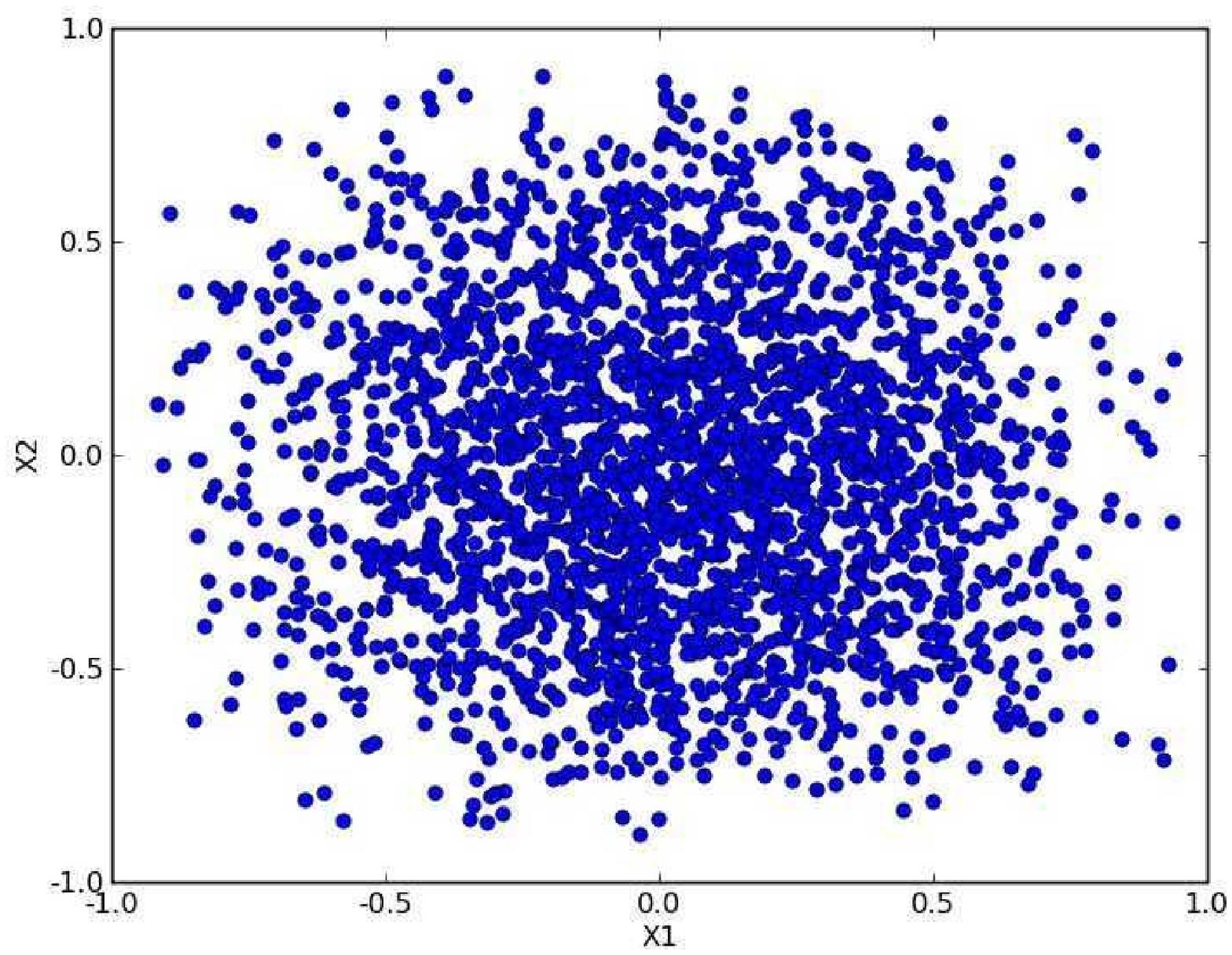}
\includegraphics[width=60mm,height=60mm]{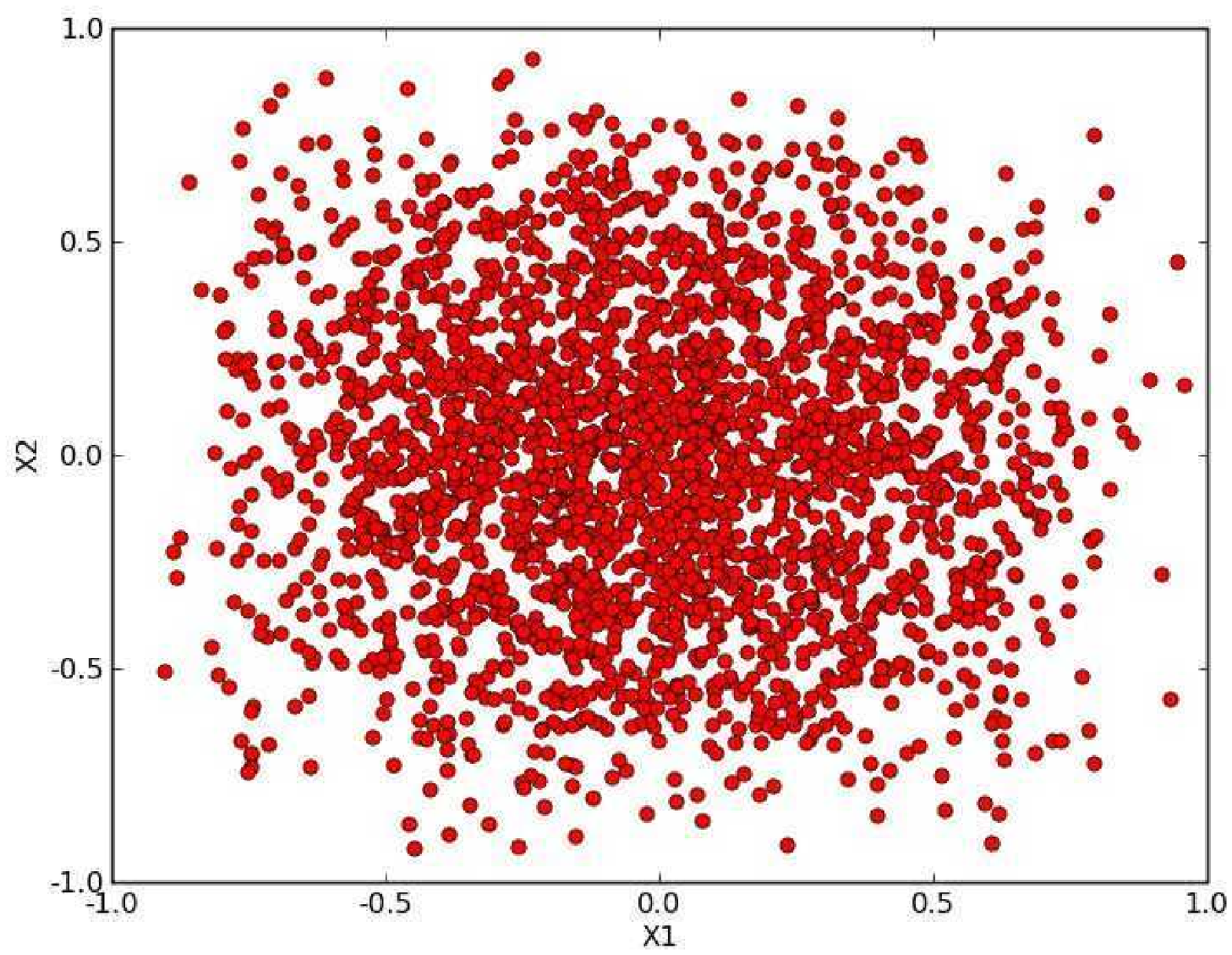}}

\noindent Figure 4. Projections to the $(x_1,y_1)$ and $(x_1,x_2)$
planes of points in phase space space generated by the chaotic
trajectory observed at intervals $\Delta t =4$ (on the left) and
randomly according the probability distribution \eqref{fourd} (on
the right).
\\

\centerline{
\begin{tabular}{ c | c | c}
Moments & Dynamics & Monte Carlo \\ \hline\hline
$<x_1^2>$ & $0.1285$ & $0.1290$  \\
\hline $
(<x_1^4>-3<x_1^2>^2)/<x_1^2>^2$ & $-0.6326$ & $-0.6098$ \\
\hline $(<x_1^2y_1^2>-<x_1^2><y_1^2>)/<x_1^2><y_1^2>$ &
$-0.05425$ & $-0.05788$ \\
\hline $<x_1^6y_1^6>$ & $0.0001276$ & $0.0001287$ \\
\hline $<x_1^{12}y_1^8>$ & $3.283\times 10^{-6}$ & $3.277\times 10^{-6}$ \\
\hline $<x_1^2x_2^2>-<x_1^2><x_2^2>$ & $-0.0001122$ & $-0.0001387$  \\
\hline $<y_1^2y_2^2>-<y_1^2><y_2^2>$ & $-0.0001884$ & $-0.0001958$ \\
\hline $\vdots$ & $\vdots$ & $\vdots$\\
\end{tabular}
} \vskip20pt

\noindent Table 2.  The entries in the column labeled `dynamics'
were obtained by numerical simulation of the dynamical system. With
the exception of the last two rows, the time duration was $20000$,
with data sampled at intervals $\Delta t=2$, starting from the
initial conditions $(x_1,y_1,x_2,y_2)=(0.1,0.2,0.11,0.09)$. The last
two rows were obtained by sampling at intervals $\Delta t=4$ over a
duration $400000$. The column labeled `Monte Carlo' was obtained by
hit and miss Monte Carlo computation using the proposed exact
distribution with 100000 accepted data points, for all but the last
two rows. The last two rows were obtained with 400000 accepted data
points.

 \vskip15pt

\section{Non-polynomial distributions}

In the previous section we considered invariant distributions which
were polynomial, with real zeroes forming a compact manifold. The
inverse method may be readily generalized to more complicated
analytic structures.

For example, consider distributions of the form,
\begin{align}\label{expdist}
\rho = Pe^{-Q}\, ,
\end{align}
where $P$ and $Q$ are polynomial\footnote{The path integral measure
in quantum field theories is often of this type. Note however that
the domain of support may be restricted (for the ergodic measure) in
a way which does not occur in quantum field theory.}. To insure that
$v$ is polynomial, we now choose an $N-2$ form with the structure
\begin{align}
\Omega = \xi e^{+Q}\, ,
\end{align}
where $\xi$ is a polynomial $N-2$ form.  Then equation
\eqref{velocity} becomes \begin{align} \label{expvel}{}^*v=
Pd\xi+2dP\wedge\xi-PdQ\wedge\xi\, .\end{align}  We have found
numerous examples of dissipative chaotic systems of this type, for
which the real zeros of P do not form a closed manifold.

As an example with $N=3$, consider
\begin{align}
P &= (1-x^2-z^6)\, , \nonumber\\
Q &= x^4+y^4+z^2+3xyz\, , \label{nonp} \\
\xi &= xyzdy+y^2dz\, .\nonumber
\end{align}
Planar projections of points generated by the resulting dynamical
system \eqref{expvel} (observed at sufficiently long intervals) and
by random sampling according to the probability distribution
\eqref{expdist} appear consistent (see figure 5) upon restricting
the domain of support of the probability distribution to $y>0$ and
$z>0$. \vskip20pt

\centerline{\includegraphics[width=60mm,height=60mm]{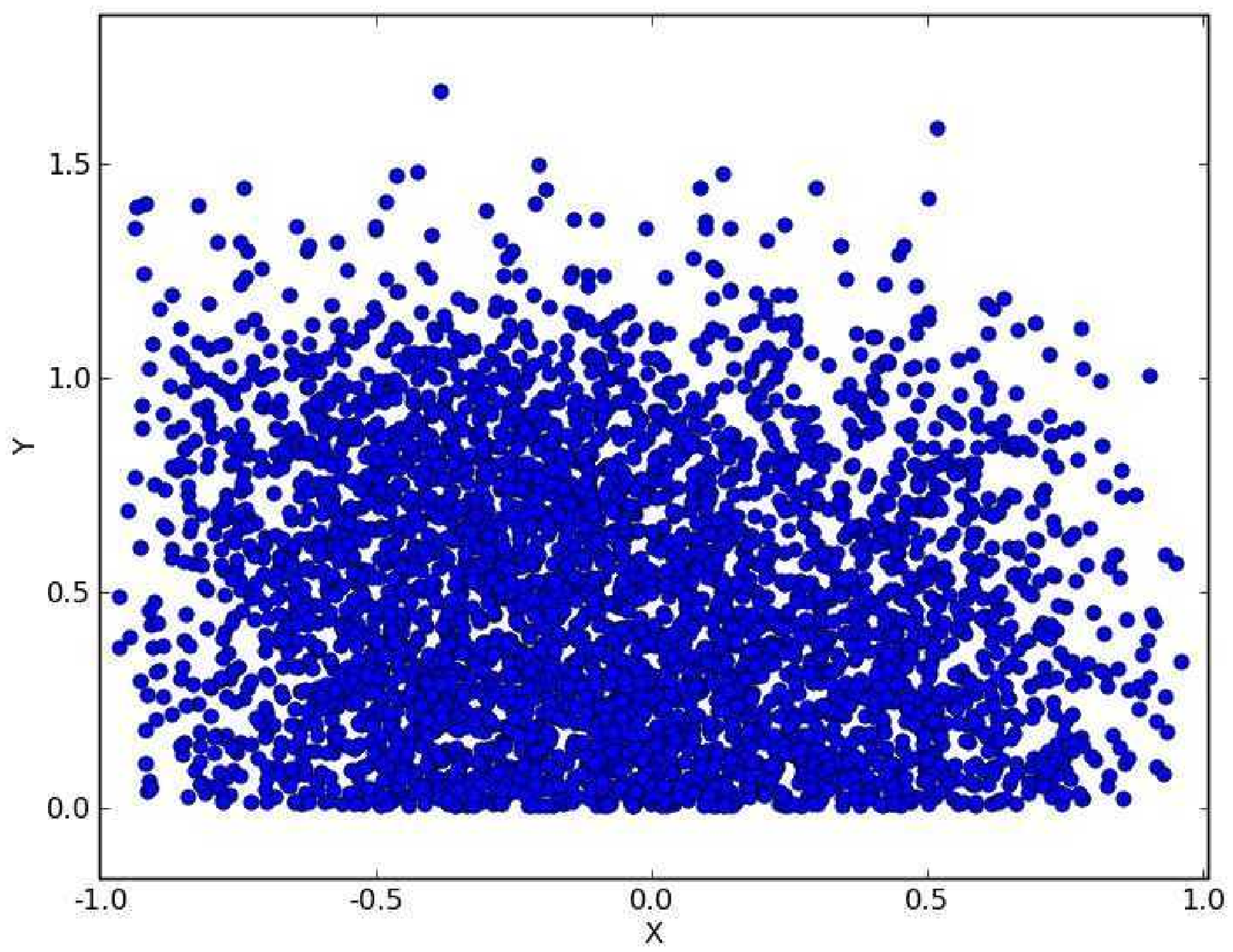},
\includegraphics[width=60mm,height=60mm]{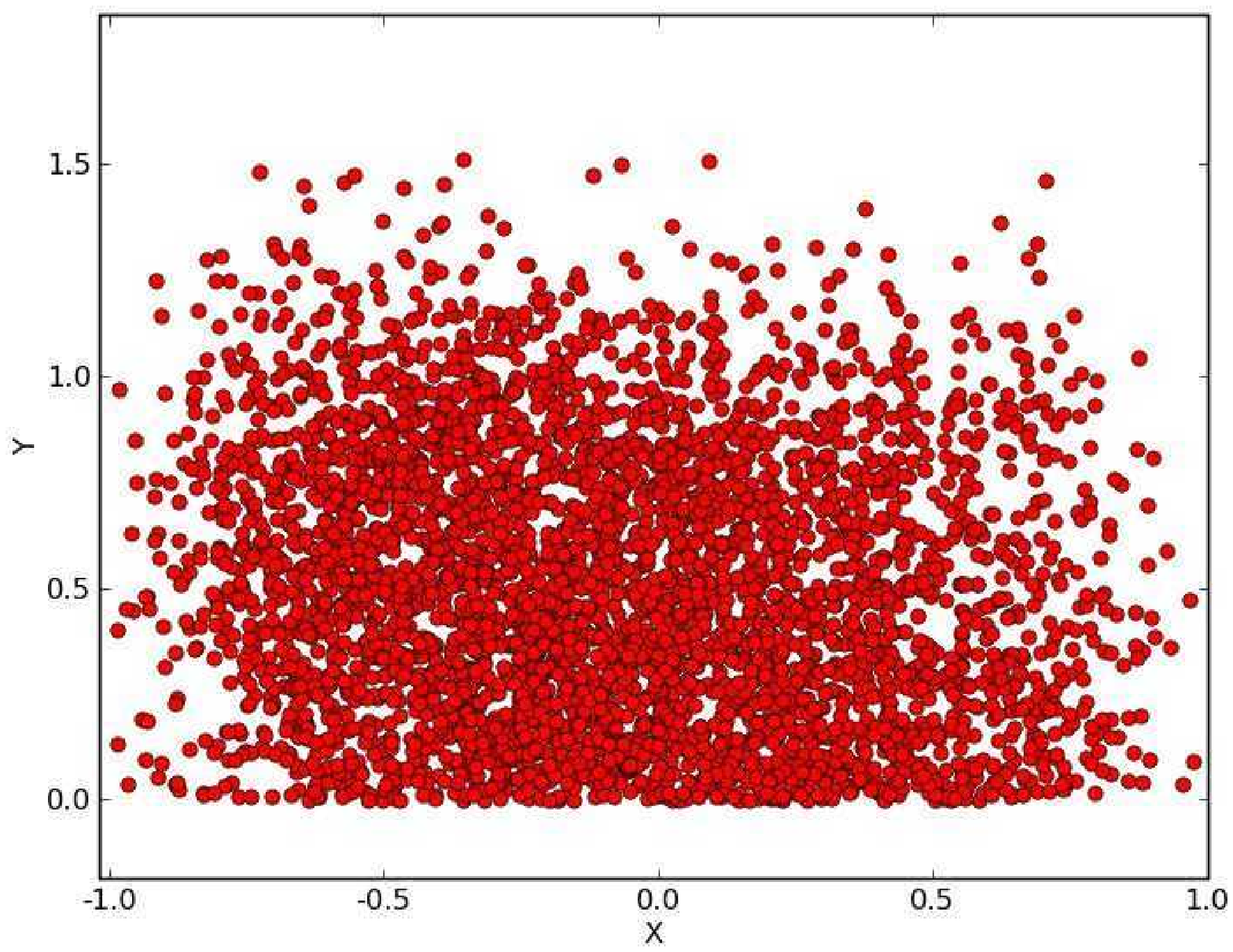}}

\centerline{\includegraphics[width=60mm,height=60mm]{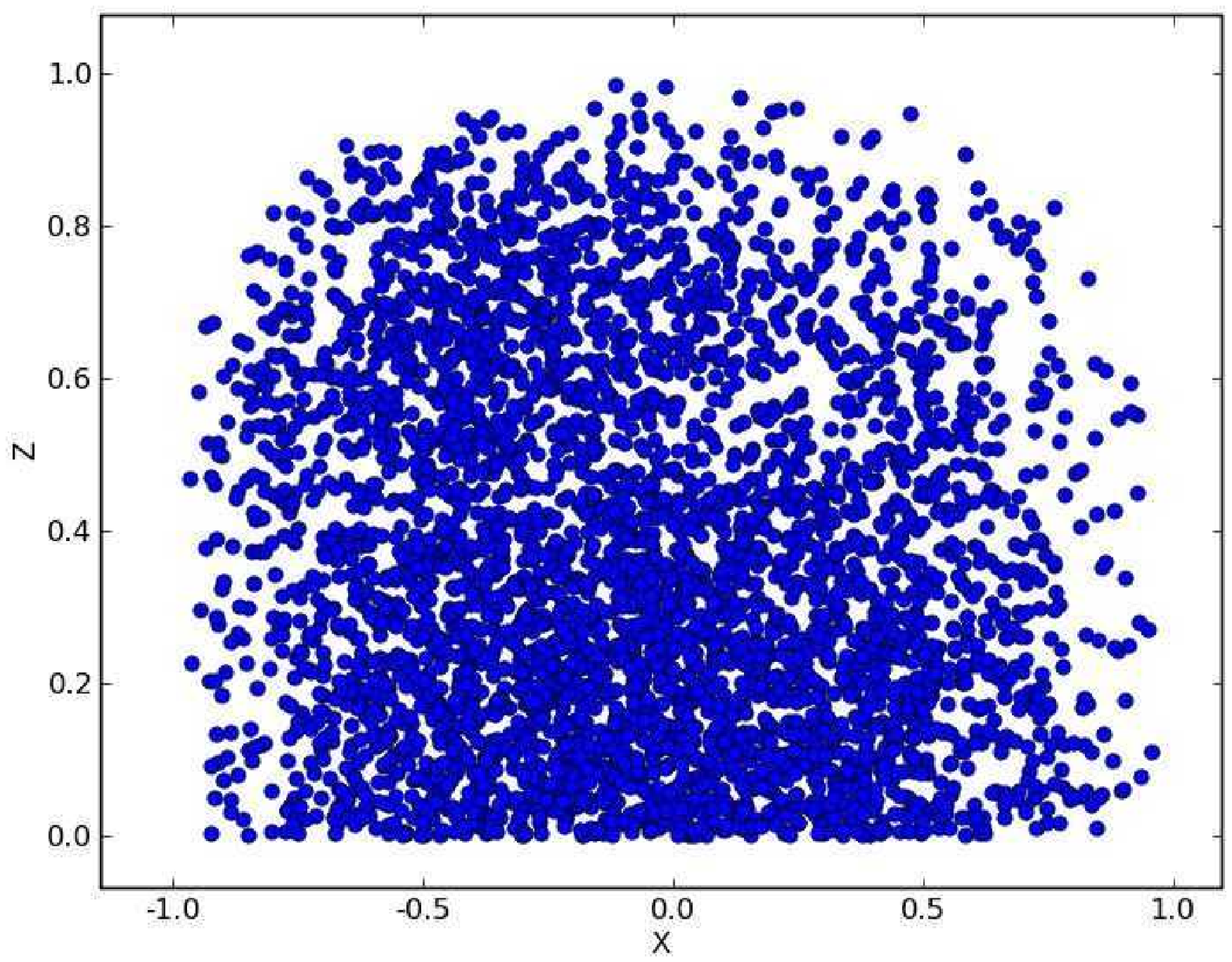},
\includegraphics[width=60mm,height=60mm]{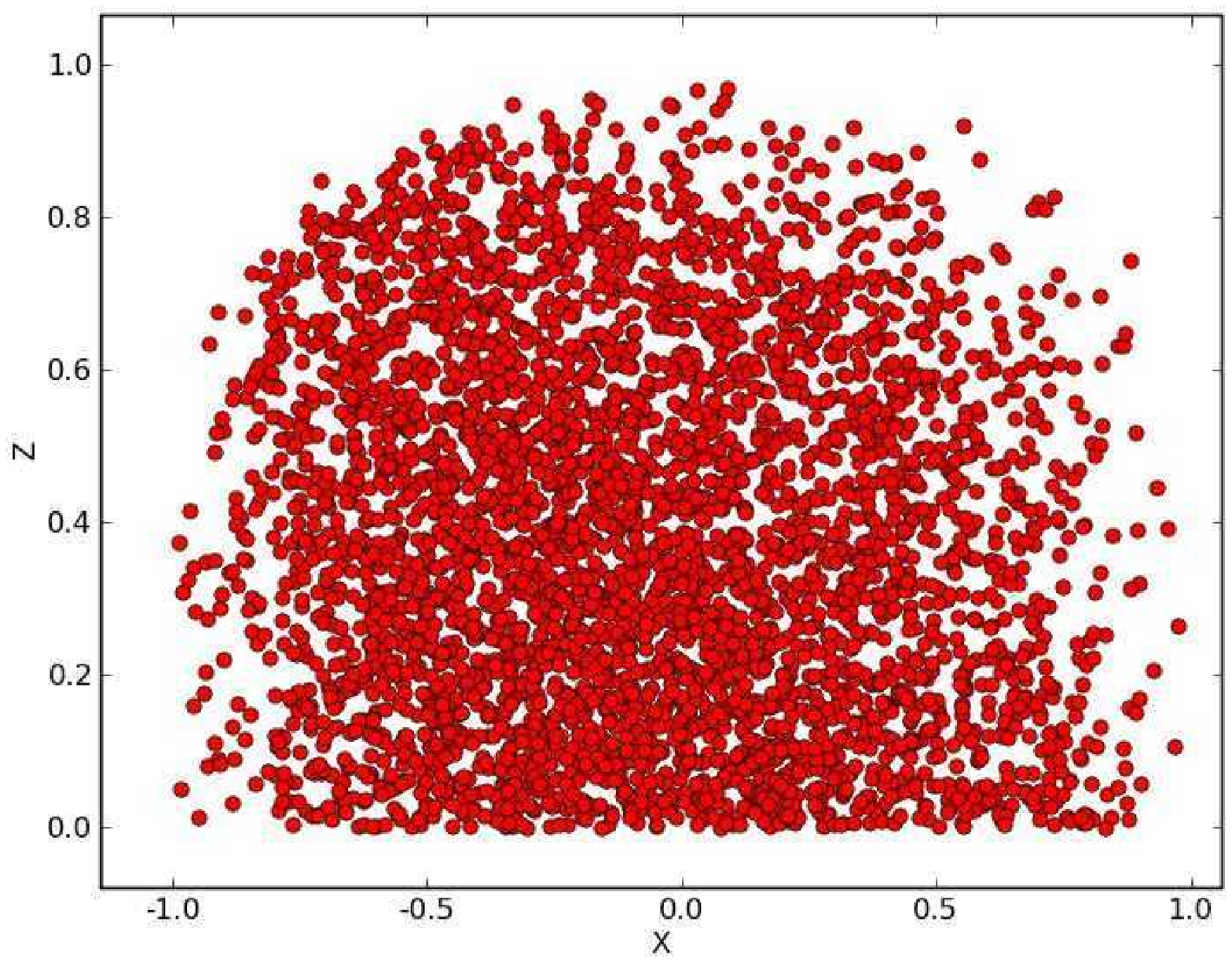}}

\noindent Figure 5. The XY and XZ projection of points (on the left)
generated by the dynamical system defined by \eqref{nonp}, observed
at time intervals $\Delta t=1$  starting from the initial condition
$(x,y,z)=(0.1,1.1,0.4)$,  and points (on the right) generated by
random sampling with the probability distribution \eqref{expdist1}.
\vskip20pt

\noindent Comparing connected moments\footnote{$<xy>_c\equiv <xy>-<x><y>$,\\
\indent\hskip8pt $<xyz>_c \equiv
<xyz>-<xy>_c<z>-<xz>_c<y>-<yz>_c<x>-<x><y><z>$, etc...} associated
with the probability distribution
\begin{align}\label{expdist1}
\rho &= (1-x^2-z^6)e^{-(x^4+y^4+z^2+3xyz)}\,\,\,{\rm for}\,\,\,
x^2+z^6\le
1,\,\,\, y>0,\,\,\,z>0\nonumber\\
\rho &= 0\,\,\,{\rm elsewhere}.
\end{align} with moments
generated by the chaotic trajectory suggests that the two agree.
(see table 3).

\vskip20pt \centerline{
\begin{tabular}{ c | c | c}
Moments & Dynamics & Monte Carlo \\ \hline\hline $<x>$ & $-0.08497$
& $-0.09152$ \\ \hline $<y>$ & $0.4898$ & $0.5046$
\\ \hline $<z>$ & $0.3712$& $0.3813$ \\ \hline
$<x^2>_c$ & $0.1697$ & $0.1685$ \\ \hline $<y^2>_c$ & $0.1052$ & $0.1028$ \\
\hline $<z^2>_c$ & $0.06145$ & $0.05809$ \\
\hline $<xy>_c$ & $-0.01830$ & $-0.01887$ \\
\hline $<xz>_c$ & $-0.01304$ & $-0.01221$ \\
\hline $<yz>_c$ & $0.001292$ & $0.0033309$ \\
\hline $<x^3>_c/(<x^2>_c)^{3/2}$ & $0.1953$ & $0.1946$ \\
\hline $<y^3>_c/(<y^2>_c)^{3/2}$ & $0.4361$ & $0.4119$ \\
\hline $<z^3>_c/(<z^2>_c)^{3/2}$ & $0.3828$ & $0.3027$ \\
\hline $<xyz>_c/\sqrt{<x^2>_c<y^2>_c<z^2>_c}$ & $-0.07400$ & $-0.07370$ \\
\hline $<x^4>_c/<x^2>_c^2$ & $-0.7019$ & $-0.7433$ \\
\hline $\vdots$ & $\vdots$ & $\vdots$ \\
\end{tabular}
} \vskip20pt

\noindent \small{Table 3.  Entries in the column labeled `dynamics'
were obtained by mumerical simulation of the dynamical system over a
time duration $40000$ , sampling at intervals $\Delta t= 0.1$,
starting from the initial conditions $(x,y,z)=(0.1,1.1,0.4)$. The
column labeled `Monte Carlo' was obtained by a hit and miss Monte
Carlo calculation using the probability distribution
\eqref{expdist1}
with 400000 accepted data points.}\\

Note that the distribution \eqref{expdist1} seems to accurately
reproduce the connected moment $<\nobreak xyz>_c$ of the chaotic
trajectory, which would not be expected if \eqref{expdist1} only
described the statistics of the trajectory upon planar projection.
This could mean that the dimension of the chaotic invariant set,
which we have not yet made an effort to calculate, is very nearly
three. If not, then perhaps \eqref{expdist1} has an interpretation
as the Fourier transform of the the Hopf functional
\begin{align}
\Psi(\vec j) \equiv \lim_{T\rightarrow\infty} \frac{1}{T} \int_0^T\,
dt\, \exp(i\vec j\cdot\vec x(t))\, , \end{align} and therefore
determines all polynomial moments $<x^ny^mz^l>$, although it must
fail to give some other expectation values\footnote{One can always
describe a distribution with support over a d-dimensional subspace
of an $N$ dimensional space in terms of a distribution $\rho$ with
support in $N$ dimensions together with a projection. However $\rho$
will only accurately determine expectation values of a subset of
functions $f(\vec x)$ on the $N$ dimensional phase space.}. This
assumes that the Hopf functional is sufficiently well behaved that
its Fourier transform has support in $N$ dimensions.

This particular chaotic system behaves both as a repeller and
attractor, depending on the initial conditions.  trajectories
outside with $x^2+z^6>1$ can never enter the chaotic invariant set.
On the other hand initial conditions with $x^2+z^6<1$ and large $z$
are rapidly drawn into the region depicted in figure 5.

Although we will not give explicit examples here, the inverse
approach can also be applied to distributions with significantly
more complicated analytic structure. For instance, one can take a
distribution of the form
\begin{align}\label{complx}
\rho = \frac{P_1}{P_2}e^{(-P_3/P_4)}\, ,
\end{align}
where $P_i$ are polynomials, together with the $N-2$ form
\begin{align}
{\cal A} = P_1^2P_4^2e^{(-P_3/P_4)}\xi\, ,
\end{align}
where $\xi$ is a polynomial $N-2$ form.  Then the velocity field
$v={}^*d{\cal A}/\rho$ will be polynomial.  One must then verify
that the distribution (or projected distribution), with some
restricted domain of support, is ergodic.  Although \eqref{complx}
may have real poles and essential singularities,  these present no
problem if they are integrable or lie outside the domain of support
of the ergodic distribution.

\section{A remark about non-uniqueness of $\rho$ and conserved quantities }

Suppose that there is a variation $\rho \rightarrow \rho+\delta\rho,
{\cal A} \rightarrow {\cal A}+ \delta{\cal A}$ under which the
dynamical system $v= {}^*d{\cal A}/\rho$ is invariant.  This
requires that
\begin{align}\label{deq}
-\frac{\delta \rho}{\rho}d{\cal A}  + d(\delta {\cal A}) = 0\, .
\end{align}
Taking the exterior derivative of this expression gives
\begin{align}
d(\frac{\delta\rho}{\rho})\wedge d{\cal A}=0\, ,
\end{align}
or equivalently
\begin{align}
d(\frac{\delta\rho}{\rho})\wedge {}^*v =0\, .
\end{align}
The last equation implies that $\delta\rho/\rho$ is a conserved
quantity.

Note that in two dimensions, ${\cal A}$ is a zero-form and
\eqref{deq} always has a solution:
\begin{align}
\frac{\delta\rho}{\rho}&= \epsilon{\cal A}\, , \nonumber\\
\delta{\cal A}&= \frac{1}{2}\epsilon{\cal A}^2\, ,
\end{align}
where $\epsilon$ is an infinitesimal constant. Since $\int
\delta\rho$ must vanish, ${\cal A}$ must have zero mean with respect
to $\rho$.  This can always be arranged by taking ${\cal
A}\rightarrow {\cal A} - <{\cal A}>$, which does not change the
dynamical system.  The existence of this solution is a reflection of
the fact that there can be no chaos with just two degrees of
freedom. Any non-trivial invariant distribution $\rho$ is a
distribution over closed orbits on which ${\cal A}$ is constant, but
does not describe the statistics of a system with a given initial
condition.

In some instances (in any dimension), the invariant probability
distribution $\rho$ used to reverse engineer a dynamical system will
not characterize individual trajectories because of the existence of
conserved quantities. This does not necessarily exclude chaos,
unless the number of conserved quantities is too large. The
distribution $\rho$ can then be interpreted as a product of
distribution over conserved quantities (which can be arbitrary) and
a distribution which is dependent on the dynamics.

As a word of caution,  not all conserved quantities associated with
non-uniqueness of the invariant distribution $\rho$ impose strong
constraints on trajectories.  Suppose for example that a dynamical
system admits two invariant distributions $\rho_1$ and $\rho_2$ with
different non-overlapping but contiguous domains of support. Then
there is a class of non-ergodic distributions of the form $x\rho_1 +
(1-x)\rho_2$ where $0\le x\le 1$.  The associated conserved quantity
is $(\rho_1-\rho_2)/(\rho_1+\rho_2)$,  which is $1$ in the region of
support of $\rho_1$ and $-1$ in the region of support of $\rho_2$.
The only constraint this conserved quantity places on trajectories
is that they can not go from the interior of region 1 to the
interior of region 2, or vice versa.

\section{Exponential distributions with Gaussian random forcing and Chaotic/Stochastic quantization}

Consider the static Fokker-Planck equation for  a dynamical system
in the presence of a Gaussian random force;
\begin{align}
d{}^*(\rho v -  \Gamma d\rho) = 0\, .
\end{align}
One can always (locally) write
\begin{align}
{}^*(\rho v -  \Gamma d\rho) = d{\cal A}\, ,
\end{align}
for some $N-2$ form ${\cal A}$.  Equivalently;
\begin{align}\label{vexp}
v = \frac{1}{\rho}({}^*d{\cal A}+\Gamma d\rho)\, .
\end{align}
Assuming $\rho = \exp(-Q)$  and ${\cal A} = \xi \exp(-Q)$, where $Q$
is a polynomial and $\xi$ is a polynomial $N-2$ form, \eqref{vexp}
becomes
\begin{align}
v = {}^*(d\xi - dQ\wedge \xi)  - \Gamma dQ\, .
\end{align}
The case $\xi=0$ is well known to quantum field theorists and arises
in the context of stochastic quantization of a field theory with
action $Q$ \cite{ParisiWu}.  The dynamical system in this case has
fixed points where $dQ=0$ with quantum fluctuations arising from
Gaussian noise. Allowing non-trivial choices for $\xi$ gives a
infinite class of possible quantizations. In the deterministic case,
$\Gamma=0$, one must take care that $\xi$ is chosen such that the
the dynamical system is chaotic.

As a highly speculative remark intended for quantum field theorists,
we suggest that there may be an advantage to considering
deterministic chaotic quantizations for which the distribution
$\exp(-Q)$ is not ergodic.  It has been noted earlier, in section 3,
that the initial distribution $\rho$ used in the inverse method may
have symmetries which are not reflected by the dynamics. Phenomena
in quantum field theory such as spontaneous symmetry breaking might
be naturally encoded in the two-form ${}^*\xi$, whereas stochastic
quantization (with $\xi=0$) does not naturally give rise to symmetry
breaking (see \cite{GP,ZG} for potentially related discussions).  If
$\exp(-Q)$ is not ergodic, one must make sure that the Schwinger
action principle equations which define quantum field theories are
still satisfied.

\section{Conclusions}

We have demonstrated an inverse method to construct chaotic
dynamical systems starting from analytic expressions for an
invariant probability distribution and a two-form.  In principle,
the inverse method can be used to generate an unlimited number of
chaotic systems together with very well motivated conjectures for
their exact statistics.

At present, the examples of dynamical systems we have considered are
not physically motivated and in some cases are very complex, having
a large number of terms in the components of the velocity field. The
next challenge is to attempt to reverse engineer systems close to
ones of physical interest.  It would ultimately be very interesting
to apply the inverse method to physical systems with a very large
number of degrees of freedom, such as fluids. The most difficult
part of reverse engineering a physical system seems to lie in
choosing the two-form, which has no direct physical interpretation.
Choosing a probability distribution is easier; one could, for
instance, choose a distribution resembling that of a fluid with some
mean shear.

Thus far, we have studied systems with only three or four degrees of
freedom. Equations of the form \eqref{velocity} can be evaluated
very quickly by hand or by computer, depending on the system, even
for very large numbers of degrees of freedom. Potential
computational problems arise only when trying to determine the
domain of support of the ergodic distribution.  At present,  we do
not have a way to do this besides direct numerical simulation of the
dynamical system,  which requires more computer time.  However,
allowing for this computer time, one should also be able to make
very well motivated conjectures about the exact statistics of
chaotic dynamical systems with a very large number of degrees of
freedom.

Given an exact invariant distribution $\rho$,  there are novel ways
to compute time-series power spectra which will be discussed
elsewhere.  In particular, one can compute the power spectra by
Monte-Carlo methods which are readily parallelizable. Moreover, it
should also be possible to give analytic expressions for the
asymptotic behavior of the power spectrum at high frequency.

While we have focused on developing an inverse method for dynamical
systems on ${\mathbb R}^N$ with polynomial velocity fields, it may
also be possible to consider more complicated phase space topologies
such as ${\mathbb S}^N \otimes{\mathbb R}^N$, which are relevant to
classical spin systems. It might also be interesting to reverse
engineer dynamical systems on the N-simplex, which arises in the
context of evolutionary biology.

We wish to emphasize that even un-physical chaotic systems and their
exact statistics, constructed by the inverse method, have an
important use. Such systems can provide a testing ground for
theoretical ideas. Moreover they can serve as benchmarks upon which
various approximation methods, such as cumulant expansions and
closure schemes for equations satisfied by mean quantities, can be
tested.

\section{Acknowledgements}

I wish to thank Tibor Antal, Richard Brower, Stephen Libby, Brad
Marston and Hisashi Ohtsuki for enlightening discussions.  I would
also like to thank the MIT center for theoretical physics for
hospitality during completion of this work.

\end{document}